\documentclass{article}

\usepackage{microtype}
\usepackage{graphicx}

\usepackage{booktabs} 

\usepackage{hyperref}
\usepackage{siunitx}

\usepackage[accepted]{arxiv}
\usepackage[skip=0.1in]{caption}
\DeclareCaptionFont{ninept}{\fontsize{9pt}{10pt}\selectfont}
\usepackage{amsmath}
\usepackage{amssymb}
\usepackage{mathtools}
\usepackage{amsthm}

\usepackage{graphicx}
\usepackage{caption}
\usepackage{subcaption}

\usepackage{xcolor}
\usepackage{color}
\definecolor{mygreen}{rgb}{0,0.6,0}
\definecolor{mygray}{rgb}{0.85,0.85,0.85}
\definecolor{mymauve}{rgb}{0.58,0,0.82}
\definecolor{mygray}{gray}{0.7}

\usepackage{listings}
\lstdefinelanguage{JavaScript}{
  keywords={typeof, new, true, false, catch, function, return, null, catch, switch, var, if, in, while, do, else, case, break, const},
  keywordstyle=\color{blue},
  ndkeywords={class, export, boolean, throw, implements, import, this},
  ndkeywordstyle=\color{darkgray}\bfseries,
  identifierstyle=\color{black},
  sensitive=false,
  comment=[l]{//},
  morecomment=[s]{/*}{*/},
  commentstyle=\color{mygreen}\ttfamily,
  stringstyle=\color{red}\ttfamily,
  morestring=[b]',
  morestring=[b]"
}

\usepackage[capitalize,noabbrev]{cleveref}

\usepackage{xspace}
\usepackage{pifont}
\theoremstyle{plain}

\theoremstyle{definition}

\theoremstyle{remark}

\usepackage[textsize=tiny]{todonotes}

\usepackage{color}
\usepackage{xcolor}

\usepackage{soul}
\usepackage{comment}
\usepackage{enumitem}
\usepackage{multirow}

\newcommand{\fullname}{UniTSyn\xspace}
\icmltitlerunning{\fullname: A Large-Scale Dataset Capable of Enhancing the Prowess of Large Language Models for Program Testing}

\begin{document}

\twocolumn[
\icmltitle{\fullname: A Large-Scale Dataset Capable of Enhancing the Prowess of Large Language Models for Program Testing}



\icmlsetsymbol{equal}{*}

\begin{icmlauthorlist}
\icmlauthor{Yifeng He}{ucdavis}
\icmlauthor{Jiabo Huang}{tencent}
\icmlauthor{Yuyang Rong}{ucdavis}
\icmlauthor{Yiwen Guo}{independent}
\icmlauthor{Ethan Wang}{ucdavis}
\icmlauthor{Hao Chen}{ucdavis}
\end{icmlauthorlist}

\icmlaffiliation{ucdavis}{UC Davis}
\icmlaffiliation{tencent}{Tencent Security Big Data Lab}
\icmlaffiliation{independent}{Independent Researcher}

\icmlcorrespondingauthor{Yiwen Guo}{guoyiwen89@gmail.com}

\icmlkeywords{Machine Learning, ICML}

\vskip 0.3in
]



\printAffiliationsAndNotice{}  

\begin{abstract}
The remarkable capability of large language models (LLMs) in 
generating high-quality code has drawn increasing attention 
in the software testing community.
However, existing code LLMs often demonstrate unsatisfactory capabilities in generating accurate and complete tests
since they were trained on code snippets collected without 
differentiating between code for testing purposes and other code.
In this paper, we present a large-scale dataset \fullname, which is capable of enhancing the prowess of LLMs for \textbf{Uni}t \textbf{T}est \textbf{Syn}thesis. 
Associating tests with the tested functions is crucial for LLMs to infer the expected behavior and the logic paths to be verified.
By leveraging Language Server Protocol, \fullname achieves the challenging goal of collecting focal-test pairs without per-project execution setups or per-language heuristics that tend to be fragile and difficult to scale.
It contains 2.7 million focal-test pairs across five mainstream programming languages, 
making it possible to be utilized for enhancing the test generation ability of LLMs.
The details of \fullname can be found in \autoref{tab:dataset_size_stat}.
Our experiments demonstrate that, 
by building an autoregressive model based on \fullname,
we can achieve significant benefits in learning and understanding unit test representations, 
resulting in improved generation accuracy and code coverage 
across all evaluated programming languages.
Code and data will be publicly available.
\end{abstract}

\section{Introduction}
\label{submission}

\begin{figure*}[ht]
    \centering
    \includegraphics[width=\textwidth]{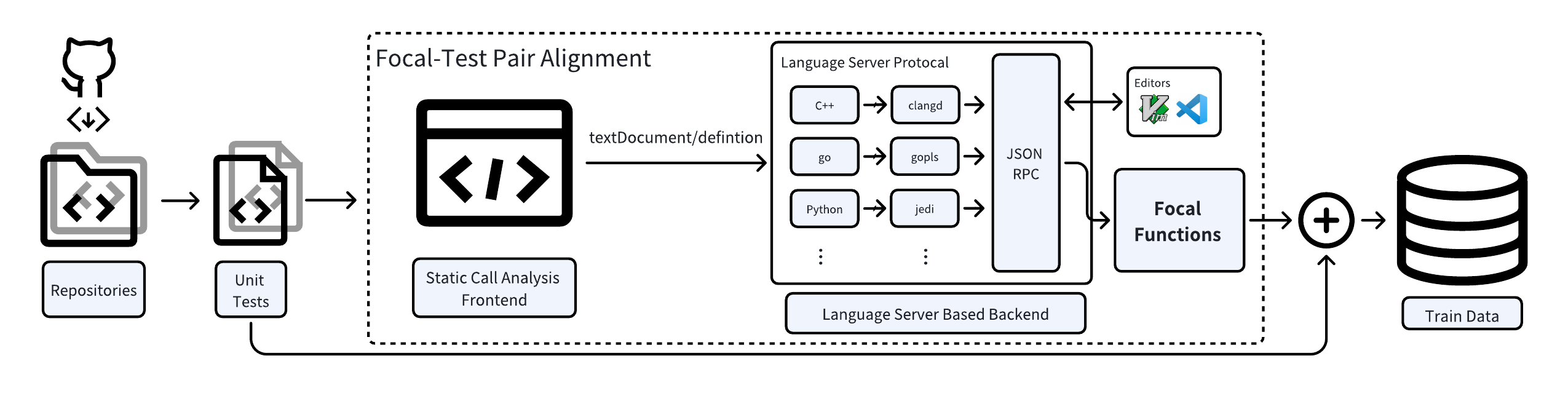}
    \caption{\fullname workflow overview. We download repositories via 
    the GitHub API and extract their unit tests. The test functions are statically analyzed to 
    identify the call to their focal functions. 
    \fullname sends JSON remote procedure 
    calls (RPCs) to language servers to get the location of the focal function definition. 
    Finally, we store the aligned function-level focal-test pairs as training data.
    }
    \vspace{-0.5\baselineskip}
    \label{fig:workflow}
\end{figure*}

Software testing is a crucial yet labor-intensive part of the software development process~\cite{zhao2017ci, beller2015when},
whose importance on early detection of program defects has been established for decades~\cite{dijkstra1972notes}.
Recently, machine learning models, especially large language models (LLMs), have shown 
their prowess in composing high-quality code~\cite{li2023starcoder,rozière2023codellama}, 
which has further fueled the software testing community's interest ~\cite{schäfer2023empirical,rao2023catlm} in applying LLMs to unit test generation or other program testing applications. 

Test generation is a more challenging task than conventional programming synthesis,
as the model not only needs executable code snippets but also precise predictions of input-output values.
Equipping off-the-shelf code LLMs~\cite{wang2023codet5p,zheng2023codegeex,nijkamp2023codegen2} 
with prompts or instructions for test generation
is poorly suited,
as demonstrated by \citeauthor{rao2023catlm}
There have been a few recent projects~\cite{nie2023TeCo, rao2023catlm}
targeting training testing-specific LLMs on code corpus that is closely related to testing.
A desired training set for this aim should include 
a large number of test functions paired with their tested source functions 
(also called \emph{focal functions} by \citeauthor{tufano2021AthenaTest} and \citeauthor{schäfer2023empirical}).

However, 
there are significant challenges surrounding the automation of collecting focal-test pairs since real-world projects do not have to follow a consistent structure.
Existing efforts
either rely on dynamic analysis or heuristics to locate focal functions~\cite{dinella2022TOGA,nie2023TeCo,alagarsamy2023a3test,tufano2021AthenaTest}
or find the coarse-grained correspondence between tests and focal functions at the file level~\cite{rao2023catlm}. 
The former category 
is \emph{difficult to scale} across programming languages, which inhibits the development of universal testing models. 
On the other hand, the latter category is limited by \emph{weak focal-test correspondences},
which hamper the models' capability to properly comprehend the expected behavior and logic paths being verified. 
These challenges underscore the need for more effective, scalable, and language-agnostic approaches 
to collect pairwise focal-test data
in an effort to fully unleash the potential of LLMs on software testing.

We present \fullname in this paper as a multilingual dataset capable of enhancing LLMs for \textbf{Uni}t \textbf{T}est \textbf{Syn}thesis.
As shown in \autoref{fig:workflow},
we integrated the Language Server Protocol (LSP) into the dataset-building process to harness its language extensibility and call-definition matching ability. 
This substantially eases the difficulty of implementing dependency analysis heuristics for different languages and executing different projects for dynamic analysis. 
Moreover, we designed a flexible, unified static analyzer to find calls to focal functions from the unit tests,
which decreases the complexity of performing call analysis for each language.

\begin{table}[t]
\setlength{\tabcolsep}{0.4cm}
    \begin{center}
        \caption{Dataset statistics.
        The ``Framework'' column indicates the static analysis we based on for test extraction.
        \#proj is the number of projects we found on GitHub for each language,
        and \#pairs is the number of focal-test pairs we collected for each language.
        }
        \label{tab:dataset_size_stat}
        \resizebox{\linewidth}{!}{
            \footnotesize
            \begin{tabular}{llrr}
                \toprule
                Language    & Framework   & \#proj & \#pairs  \\
                \midrule
                Python      & unittest,pytest           & 43,848 & 1,218,311  \\
                Java        & JUnit           & 25,488  & 1,097,518  \\
                Go          & testing           & 38,097  & 361,075    \\
                C++         & GoogleTest           & 20,090  & 25,513     \\
                JavaScript  & MochaJS           & 17,621  & 13,293     \\
                \bottomrule
            \end{tabular}
        }
    \end{center}
\end{table}

To explore the quality of the \fullname dataset, 
we further trained an autoregressive model called UniTester to 
synthesize tests in different programming languages. 
\fullname yielded non-negligible performance advantages 
on generating accurate and complete tests
over several up-to-date LLMs intended for both code and test synthesis,
which demonstrates the importance of our proposed \fullname as a flexible dataset collection framework.

To sum up, we make the following contributions:
(1)  We present a large-scale dataset of 2.7M focal-test function pairs across five commonly used programming languages,
which will be useful in advancing the field of software engineering
through LLM coding assistance.
(2) We release our generic and easily applicable approach for building multilingual unit test datasets with function-level focal-test alignment.
Our approach is extendable to any language that has a mature implementation of LSP,
allowing the LLM to broaden its testing capabilities in more diverse 
software engineering scenarios.
(3) We validated the quality of \fullname by training an autoregressive language model on it. The more accurate and complete tests generated by our model compared to existing test and code LLMs demonstrate the necessity and benefits of training with explicit correspondence between tests and focal functions for multilingual test generation.

\section{Related Work}

\subsection{Code Understanding and Generation}

The application of Machine Learning (ML) in Software Engineering (SE) has gained significant attention recently,
particularly with the development of LLMs.
LLMs can help SE in various ways, including
code generation \cite{chen2021codex, li2022alphacode, nijkamp2022codegen, li2023starcoder, rozière2023codellama, wang2021codet5},
code summarization \cite{iyer-etal-2016-summarizing, leclair2019neural, ahmad-etal-2020-transformer},
and code classification \cite{feng-etal-2020-codebert, guo2022unixcoder, zhao2023understanding, huang2023code}.
To facilitate the study of ML for SE, a variety of datasets have been built.
These datasets are either collected from
competitive programming contests, like POJ104 in the CodeXGlue benchmark \cite{lu2021codexglue}
and CodeNet \cite{puri2021codenet}, or open-source software like 
the CodeSearchNet challenge \cite{husain2020codesearchnet}
and CoderEval \cite{yu2023codereval}.
Datasets specialized for evaluating the code generation performance of LLMs like HumanEval \cite{chen2021codex}
and HumanEval-X \cite{zheng2023codegeex} have also been established using algorithmic coding problems formatted just
like on LeetCode. 
These datasets are built for general-purpose program synthesis like CodeT5+ \cite{wang2023codet5p}, CodeGen2 \cite{nijkamp2023codegen2},
InCoder \cite{fried2022incoder}, and SantaCoder \cite{allal2023santacoder}.
However, as shown by \citeauthor{zhao2023understanding} and \citeauthor{huang2023code},
test cases can greatly help improve the model's ability to understand code.
With few datasets focusing on test cases, the necessity of \fullname is justified.

\subsection{Unit Testing}

Unit testing is a common self-assessment testing technique
where developers use a set of inputs and outputs of their code  
to validate that the code is working as expected \cite{zhu1997unittest}.
In this setting, a test case consists of an input and the corresponding output after the code execution.

To evaluate the completeness and comprehensiveness of test cases,
code coverage is often used as a common metric \cite{qian2006coverage_survy, nagappan2010evidence}.
Code coverage measures the percentage of the code that is executed. 
Statement, line, and branch coverage are often used depending on the coarseness of the testing requirements.
Code coverage is measured because executing a piece of code is the necessary condition for finding bugs in it \cite{nagappan2010evidence}.
In SE, coverage-guided software testing
has also shown its power in detecting bugs in various software domains 
\cite{chen2016jvm, matryoshka,  serebryany2016libfuzzer, angora, fioraldi2020afl_pp, rong2020integrity, rong2024valkyrie},
highlighting the importance of this metric.
In general, unit testing is an indispensable part of the modern software development ecosystem,
ensuring the quality and reliability of systems across various domains.
Therefore, constructing test cases with high coverage is a necessity.

\subsection{Software Testing via Machine Learning}

The goal of test generation is to utilize ML models to aid software testing,
which can be achieved via prompting or instructing general-purpose code LLMs \cite{wang2023codet5p, zheng2023codegeex, nijkamp2023codegen2},
or training testing-specific LLMs.
ATLAS \cite{watson2020learning_assert}, AthenaTest \cite{tufano2021AthenaTest}, TOGA \cite{dinella2022TOGA}, A3Test \cite{alagarsamy2023a3test},
TeCo \cite{nie2023TeCo}, and CAT-LM \cite{rao2023catlm} are testing-specific LLMs based on the transformer architecture.
Some testing-specific LLMs are trained on large-scale test functions
and their aligned focal functions \cite{nie2023TeCo, rao2023catlm},
where focal functions are the functions being tested \cite{tufano2021AthenaTest, schäfer2023empirical}.
To align test and focal functions,
some work relies on dynamic execution context or heuristics for locating focal functions 
\cite{dinella2022TOGA,nie2023TeCo,alagarsamy2023a3test,tufano2021AthenaTest}.
The extensibility of this approach is limited, since automating the setup of dynamic execution for 
different projects is challenging even within the same language.
For example, TeCo's execution-based data collection was only applied to 1270 Java projects, 
which is one of the easiest languages for cross-platform execution thanks to the Java Virtual Machine (JVM) and the Maven build system.
The build system sets up the project in a way that makes it easy to automate the execution.
Extending this method to other languages would likely be prone to complications not present in TeCo.
On the other hand, relaxing the alignment at file-level \cite{rao2023catlm} is easier to scale up than the previous approach. 
However, this weak focal-test correspondence disrupts the models' ability to thoroughly understand the expected behavior and logic paths in the focal function.

\section{Design of \fullname Dataset}

\subsection{Challenges}

As the interest in using LLMs for test generation grows,
the inherent limitations of their underlying datasets are becoming increasingly apparent.
Models like CAT-LM \cite{rao2023catlm}, specifically designed for test generation, 
suffer from a lack of flexibility to adapt to various languages due to dataset constraints.
In contrast, LLMs trained on general-purpose code datasets like SantaCoder \cite{allal2023santacoder} 
can incorporate new languages by pulling code from platforms like GitHub.
However, their training data often lacks a crucial link between the test functions and the focal functions. 
Without this correspondence,
the models face challenges in deducing the intended behavior and logic paths of focal functions when generating tests.
Consequently, this leads to tests that are not precise or thorough.
we aim to build a dataset that emphasizes unit tests and their corresponding
focal functions while being multilingual.
This leads to two major challenges.
First, analyzing tests for their focal function calls requires domain knowledge and 
language-specific rules since different languages have distinct grammars and
unique syntax for unit tests.
Second, extracting precise focal function definitions from the dependency graph
is labor-intensive and unique to each language.
To overcome these challenges, we abstracted the differences in languages away from the
static analysis pipeline to build a generalized static call analyzer that operates
on the Abstract Syntax Tree (AST) of different languages,
and integrate the existing language servers to locate the focal function definition
via their dependency analysis.
This design not only reduces the difficulty of analyzing calls across different programming languages
but also eliminates the requirement of implementing several dependency analyses or setting up multiple execution environments.

\subsection{Data Collection}

Our dataset contains a large assemblage of data
collected from open-source software.
We used CAT-LM's \cite{rao2023catlm} approach to locate 
repositories on GitHub that are under active development. 
More specifically, we mined repositories in Python, Java, Go, C++, and JavaScript 
that have more than 10 stars and new commits after Jan 1st, 2020
to extend CodeSearchNet \cite{husain2020codesearchnet}.
In addition, we filtered out repositories that are archived, forked, or mirrored.

\subsection{Dataset Construction}
Providing more fine-grained data with multilingual focal-test alignment
is useful to machine learning for understanding the implementation of focal functions
\cite{dinella2022TOGA, nie2023TeCo, alagarsamy2023a3test, tufano2021AthenaTest}.
We designed \fullname to achieve this goal.
The frontend identifies potential test functions and 
locates the focal functions.
The backend will then retrieve the source code of the focal function from the
repository codebase.
\autoref{fig:workflow} shows an overview of the workflow for obtaining our dataset \fullname.

\begin{figure}[t]
    \centering
    \includegraphics[width=0.45\textwidth]{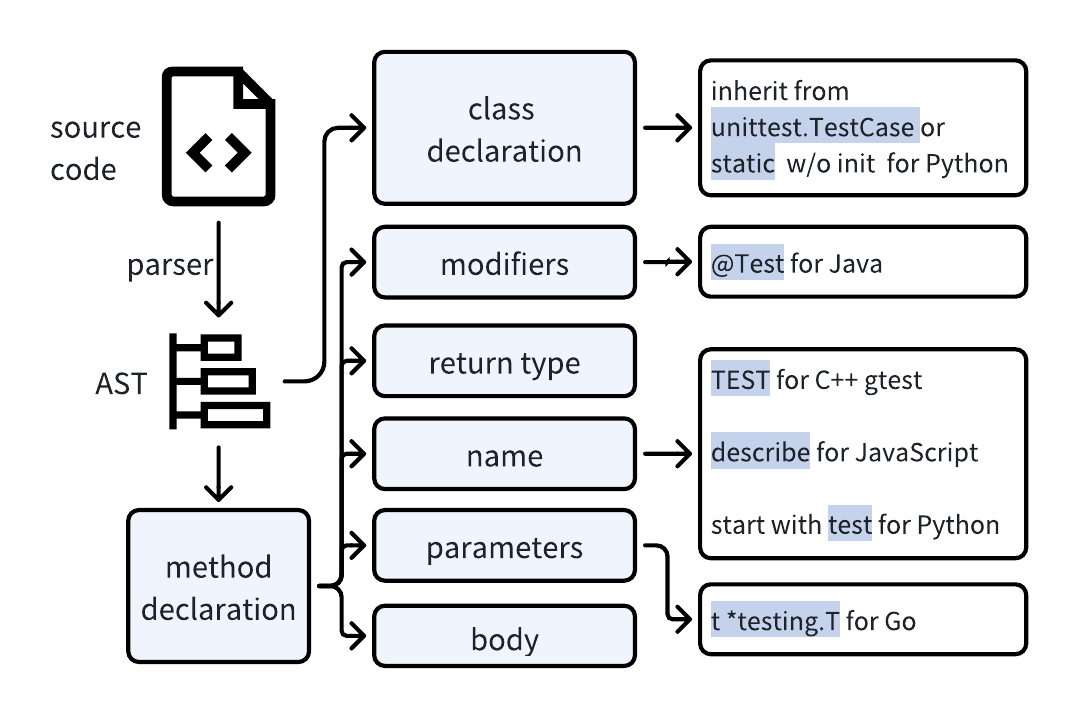}
    \caption{Test collection phase of \fullname frontend.
    The highlighted words are identifiers
    that we use to locate testing functions.}
    \label{fig:collection_phase}
\end{figure}

\paragraph{Parsing.}

To construct a multilingual unit test dataset,
we need to generalize the static analysis process among different languages.
The first step in any static analysis is to parse the source code into AST,
which can be achieved using each language's compiler or interpreter.
To ease the differences in invoking different tools for different languages,
we selected tree-sitter \cite{tree-sitter} as the backbone of our parsing process.
Tree-sitter can parse any programming language that has a formal syntax definition.
Using the tree-sitter, we designed an easily extendable parsing and AST interface to
enable smooth static analysis for all languages.

\paragraph{Test Function Identification.}

We traverse ASTs to find the test functions.
To mitigate the differences between languages, we provide an interface to determine whether an AST contains any test functions.
This interface takes hooks as call-back functions to check for test functions.
In this work, we present two implementations. 
One relies on heuristics by checking the function name. 
The other uses language-specific features to determine if a function is a test. 
For example, in Java \texttt{JUnit},
all test functions need to have the \texttt{@Test} modifier.
\autoref{fig:collection_phase} illustrates the test collection phase
and the aforementioned hooks are summarized to the right of the figure.
To extend this framework to other languages and testing suites or to improve the success rate of test function identification,
one could provide a new, specially designed hook.

\begin{figure}[t]
    \centering
    \includegraphics[width=0.45\textwidth]{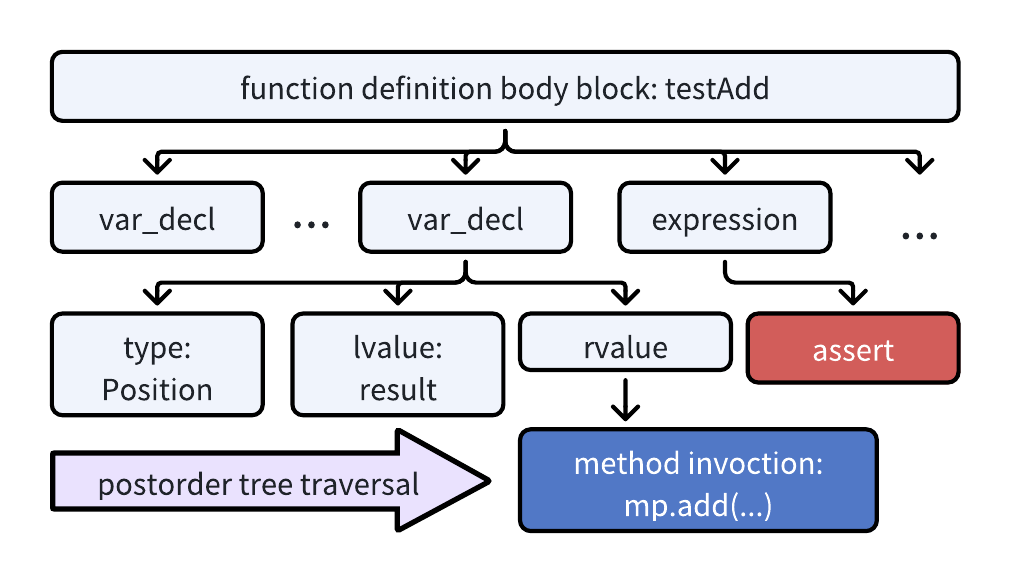}
    \caption{Focal function call analysis phase of \fullname frontend.
    This figure shows a simplified AST of the example test function in \autoref{lst:java_test_func}.
    The red node in the AST is the first encountered assertion node with a postorder 
    tree traversal.
    \texttt{var\_decl} is the abbreviation of the variable declaration nodes.}
    \label{fig:focal_analysis}
\end{figure}

\paragraph{Focal Function Call Analysis.}
Given a test function, we need to identify its focal function call.
Since a unit test function usually makes multiple function calls to set up its environment,
it is hard to identify the real call to the focal function.
In this paper, we followed TeCo's \cite{nie2023TeCo} practice to select the last function 
call that invokes a function definition in the repository
before the first assertion.
\autoref{lst:java_test_func} demonstrates an example of a Java test case,
where line \autoref{lst:java_test_func:bold} includes the focal function call.
To allow easy extension to different languages as a unified method,
we designed another interface that the developers can implement with one extra function. 
The static call analysis phase to build \fullname is demonstrated in \autoref{fig:focal_analysis}.
Our framework's generic focal-call analysis algorithm
only requires one extra function to adapt to a new language.

\begin{lstlisting} [
    belowskip=-2\baselineskip,
    float,
    language=Java, 
    caption={Example Java test function and its paired focal}, 
    label=lst:java_test_func,
    basicstyle=\ttfamily\scriptsize,
]
@Test
public void testAdd() {               
    int x = 500;     int offsetX = 100;
    int y = 700;     int offsetY = 200;
    Position mp = new Position(x, y);
    Position result = (*@\ttfamily\bfseries  mp.add(offsetX, offsetY); @*)  (*@ \label{lst:java_test_func:bold} @*)
    assertEquals(mp.x + offsetX, result.x);
}
public Position add(int x, int y) {  
    return new Position(this.x + x, this.y + y);
}
\end{lstlisting}

\paragraph{Focal Function Extraction via LSP.}

We incorporated the LSP into our dataset-building process to utilize its adaptability to different 
languages.
This  design helps relieve the pressure on dependency analysis
across various languages and executing multiple projects.
LSP is a language-neutral and standardized protocol 
for communication between language servers and editor clients \cite{LSP},
decoupling the programming languages' features and functionalities from
integrated development environments 
\cite{rask2021specification}.
This separation allows the language intelligence to run as an independent service,
and hence be integrated into our dataset construction.
We developed the \fullname backend as a client to the language servers,
where we send 
requests to the language servers to locate the definition of the focal function.
The language servers send back the location of the focal function as a response
\cite{gunasinghe2021language}.
Our approach of using LSP can easily extend to other languages as long as the language's language server is implemented.
Our proposed method of aligning function-level focal-test pairs using LSP is easily extendable,
requiring only the commands to start the corresponding language server.

\section{Experiment}

We built a test generation model called \textbf{UniTester},
which was trained on \fullname
and is capable of synthesizing unit tests for programs in different languages.
To investigate the quality of our testing code corpus,
we evaluated UniTester and several up-to-date code or test generation models on
HumanEval-X~\cite{zheng2023codegeex},
as a popular multilingual program synthesis benchmark
that shares the same programming languages involved in \fullname.

\subsection{UniTester: A Unified Test Generation Model}
We constructed our model based on established code generation practices~\cite{radford2019gpt2}.
We utilized an autoregression signal~\cite{radford2019gpt2,rozière2023codellama,nijkamp2022codegen} 
for continual training
of SantaCoder~\cite{allal2023santacoder}, which is a powerful yet lightweight 
state-of-the-art code generation model composed of 1.1B parameters. 
We selected this model for its size, as this allows
it to strike a balance between effectiveness and efficiency.
To build a training sample, 
we concatenated each focal-test function pair with a newline symbol. 
We packed the training corpus and 
sampled sequences of a constant length to feed into the models, 
thereby avoiding padding samples of varying lengths and enhancing training efficiency. 
We set the constant sequence length at $2048$ with a batch size of $32$. 
Our training process employed a learning rate of $5e^{-5}$, 
incorporating a logarithmic warmup for the initial 500 steps and a cosine annealing strategy~\cite{loshchilov2016cosine} for the rest. 
We also adopted a weight decay of $0.05$ to refrain from overfitting and catastrophic forgetting~\cite{kirkpatrick2017overcoming}. 
Our UniTester was trained by an Adam optimizer~\cite{kingma2014adam} for $36,000$ steps on the collected data containing around one billion tokens.
The entire training process using eight Nvidia V100 GPUs spanned approximately 24 hours.

\subsection{Research Questions and Evaluation Setup}

To study the contributions of our collected testing code corpus, \fullname, we investigated the following four questions:

\begin{enumerate}[label=\textbf{RQ.\arabic*}, itemindent=10pt]
\item \textbf{How accurate are the test cases generated by LLMs? }
Considering that the primary objective of software testing is to 
identify potential flaws in code, 
it is essential to ensure that the generated test cases are 
accurate to minimize 
incorrect evaluations.
This requires models to not only comprehend the general semantics of focal functions 
but also reason about their specific behavior and precise input-output mappings,
which is fundamentally challenging. 
In this case,
we followed \citeauthor{chen2022codet} to parse the assertions from the generated tests to examine their standalone correctness and derive conclusions without mutual influences.

\item \textbf{How many of the generated tests are complete?}
We define the completeness of tests in terms of both their executability 
and the proportion of code in the focal functions that has actually been executed by them.
To evaluate how complete the tests generated by LLMs are, 
we took the raw outputs of models without intricate post-processing as individual tests
and reported the line/statement and branch coverage rates
they achieved on the corresponding focal function.
This assessment allows us to better understand the effectiveness of LLM-generated tests 
in covering various conditions and scenarios within the codebase without excessive manual intervention.

\item \textbf{Is it necessary to train LLMs with pairwise focal and test functions?}
\fullname is designed to identify test functions and pair them with their targets 
in complex real-world software projects. 
Given that the composition of test functions heavily depends on 
the expected behavior of their target focal functions, 
we were motivated to train using test-focal function pairs. 
To justify our motivation, 
we broke the pairwise connections and treated test and focal functions as 
independent code snippets for LLMs training. 
The resulting models were then compared with 
our UniTester that was trained with pairwise data.

\item \textbf{What are the effects of training with multilingual testing code?}
While previous work~\cite{feng-etal-2020-codebert,guo2022unixcoder} 
have shown that training language models on 
data of different distributions
can be beneficial, 
it was unclear whether this conclusion also held for test generation. 
Therefore, we compared models solely trained on Python to UniTester
that was trained on five different languages. 
This comparison aimed to determine whether test generation models 
should be language-specific or if they can be universal across languages.

\end{enumerate}

\paragraph{Evaluation datasets.}
To address the aforementioned research questions, 
we evaluated UniTester and the state-of-the-art code LLMs on
HumanEval-X~\cite{zheng2023codegeex}, 
which is a popular and multilingual code generation benchmark dataset.
It comprises natural language descriptions of $164$ coding tasks and 
their solutions in Python, C++, Java, JavaScript, and Go
(the same languages used in \fullname).
Note that we prompted the models to generate tests 
for the given set of programs in HumanEval-X
instead of synthesizing the programs based on their problem descriptions.

To mitigate data leakage and facilitate a more straightforward execution setup,
we chose to experiment on an existing code generation benchmark.
Previous work~\cite{rao2023catlm,nie2023TeCo,schäfer2023empirical} 
on test generation typically validate their models on public data collected by themselves.
Given that the success of the latest code LLMs is 
also based on large-scale open-source coding resources, 
it is challenging to avoid data leakage in such cases. 
Consequently, we turned to the benchmark datasets intended for this purpose, 
ensuring that all focal and test functions used for assessment had never been seen by any of the models. 
HumanEval-X is also well crafted such that its test cases can be executed without excessive efforts in
setting up the execution environments and
sorting out the intricate dependencies between packages.
This allows for a more straightforward and unbiased evaluation of 
the models' performance in generating accurate and comprehensive tests for software projects.

\paragraph{Instructive prompts.}
Following CodeT's approach~\cite{chen2022codet}, 
we prompted the models to generate tests using 
language-specific assert keywords.
Examples of prompts for the first task in HumanEval-X are shown in \autoref{fig:prompts}.
Specially,
for models trained by instruction tuning,
we concatenated the focal function and our natural language hint (``\texttt{Check the correctness of...}'')
as the instruction then asked the models to complete the test functions.
The maximal input length was set to $800$ and the synthesized outputs were allowed to have another $256$ tokens at most.
We set the temperature for generation to $0.2$ following \citet{xiong2023program,rao2023catlm}
and kept it consistent for all models,
then sampled ten outputs for each task. 
Subsequently, we parsed and sampled the first ten assertions from 
the generated tests for accuracy computation on each task.
This ensures that different models produce 
a similar number of assertions for verification.
Regarding the evaluation of completeness,
we treated each output of the models as an individual test and executed it independently.
We avoided applying intricate post-processing on all of the models' outputs except for the Java ones,
on which we found that adding closing brackets for the test class is helpful to most models for ensuring executability.
The average coverage rates from ten trials are reported in \autoref{tab:humaneval_cover}.

\begin{figure}[t]
  \begin{minipage}{\linewidth}
    \begin{subfigure}[t]{\linewidth}
      \begin{lstlisting}[
        language=python, 
        caption={Python}, 
        label=lst:prompt:py,
        basicstyle=\ttfamily\scriptsize,
      ]
from typing import List
def has_close_elements(
  numbers: List[float], threshold: float) -> bool:
    ...
# Check the correctness of `has_close_elements`
def test_has_close_elements():
    assert has_close_elements(
      \end{lstlisting}
    \end{subfigure}
    \vspace{-0.4\baselineskip}
    \begin{subfigure}[t]{\linewidth}
      \begin{lstlisting}[
        language=go, 
        caption={Go}, 
        label=lst:prompt:go,
        basicstyle=\ttfamily\scriptsize,
      ]
func HasCloseElements(
  numbers []float64, threshold float64) bool 
{    ...    }
// Check the correctness of `HasCloseElements`
func TestHasCloseElements(t *testing.T) {
    assert := assert.New(t)
    assert.Equal(HasCloseElements(
      \end{lstlisting}
    \end{subfigure}
    \vspace{-0.4\baselineskip}
    \begin{subfigure}[t]{\linewidth}
      \begin{lstlisting}[
        language=JavaScript,
        caption={JavaScript}, 
        label=lst:prompt:js,
        basicstyle=\ttfamily\scriptsize,
      ]
const hasCloseElements = (numbers, threshold) => {...}
// Check the correctness of `hasCloseElements`
const testHasCloseElements = () => {
    console.assert(hasCloseElements(
      \end{lstlisting}
    \end{subfigure}
    \vspace{-0.4\baselineskip}
    \begin{subfigure}[t]{\linewidth}
      \begin{lstlisting}[
        language=C++, 
        caption={C++}, 
        label=lst:prompt:cpp,
        basicstyle=\ttfamily\scriptsize,
    ]
bool has_close_elements(
  vector<float> numbers, float threshold)
{    ...    }
// Check the correctness of `has_close_elements`
#undef NDEBUG
#include <assert.h>
int main(){
    assert(has_close_elements(
      \end{lstlisting}
    \end{subfigure}
    \vspace{-0.4\baselineskip}
    \begin{subfigure}[t]{\linewidth}
      \begin{lstlisting}[
        language=Java, 
        caption={Java}, 
        label=lst:prompt:java,
        basicstyle=\ttfamily\scriptsize,
    ]
class Solution {
    public boolean hasCloseElements(
      List<Double> numbers, double threshold) 
    {    ...    }
}
public class Main {
    // Check the correctness of `hasCloseElements`
    public static void main(String[] args) {
        Solution s = new Solution();
        assert s.hasCloseElements(
    \end{lstlisting}
    \end{subfigure}
    \vspace{-0.4\baselineskip}
  \end{minipage}
  \caption{Prompts used for test generation in different languages}
  \label{fig:prompts}
\end{figure}

\paragraph{Compared models.}
We conducted extensive comparisons between our proposed model and 
both the state-of-the-art code generation models highlighted in \citeauthor{xiong2023program}'s paper
as well as the latest test generation model~\cite{rao2023catlm}. 
In our evaluation, 
we considered code generation models with a similar number of parameters to ours, 
encompassing both encoder-decoder (CodeT5+~\cite{wang2023codet5p}) and 
decoder-only structures 
(CodeGen2~\cite{nijkamp2023codegen2}, WizardCoder~\cite{luo2023wizardcoder}, InCoder~\cite{fried2022incoder} and SantaCoder~\cite{allal2023santacoder}). 
For test generation, 
we selected CAT-LM~\cite{rao2023catlm} as our competitor, 
which was trained on unit tests from Python and Java 
projects that were paired with their target functions at the file level.
CAT-LM was chosen because 
it is trained on a much larger number of tokens 
($60$B \textit{v.s.} $1$B) 
than our model and has double our model size. 
This allowed us to demonstrate our model's superior performance, 
despite any potential advantages stemming from the amount of training data and the model size.

\subsection{Evaluation Results}

\paragraph{RQ.1 How accurate are the test cases generated by LLMs?}
\begin{table}[t]\footnotesize
\setlength{\tabcolsep}{0.05cm}
\caption{
Evaluations of the accuracy of tests generated by LLMs.
 The ``\#Params'' column indicates the size of models.
The best results are highlighted in bold.
The models marked with $\dagger$ are intended for test generation.
}
\label{tab:humaneval_acc}
\begin{tabular}{lcccccccccccccc}
\toprule
Model &  & \#Params &  & Py &  & C++ &  & Java &  & JS &  & Go &  & Avg\\ 
\midrule
CodeT5p &  & 770M &  & 30.6 &  & 33.7 &  & 26.9 &  & 37.1 &  & 32.9 &  & 32.2\\ 
CodeGen2 &  & 1.0B &  & 34.0 &  & 40.7 &  & 24.1 &  & 30.5 &  & 36.1 &  & 33.1\\ 
WizardCoder &  & 1.0B &  & 36.8 &  & 43.9 &  & 28.7 &  & 31.3 &  & 47.7 &  & 37.7\\ 
InCoder &  & 1.3B &  & 34.2 &  & 33.5 &  & 22.6 &  & 24.4 &  & 31.5 &  & 29.2\\ 
SantaCoder &  & 1.1B &  & 36.2 &  & 34.7 &  & 36.5 &  & 30.6 &  & 31.5 &  & 33.9\\ 
CAT-LM$^\dagger$ &  & 2.7B &  & 37.5 &  & 31.6 &  & 34.4 &  & 29.2 &  & 36.9 &  & 33.9\\ 
\textbf{UniTester$^\dagger$ (Ours)} &  & 1.1B &  & \textbf{52.5} &  & \textbf{55.1} &  & \textbf{48.8} &  & \textbf{41.7} &  & \textbf{59.7} & \textbf{} & \textbf{51.5}\\ 
\bottomrule
\end{tabular}

\end{table}
The accuracy of the test cases generated by different models is presented in \autoref{tab:humaneval_acc}. 
Our model achieved up to a $40\%$ relative margin on Python compared to the top competitor (CAT-LM), 
demonstrating remarkable advantages over both code and test synthesis models. 
In terms of the code generation models,
UniTester beat the strongest competitor (WizardCoder) by $36.6\%$ on average.
These results validate the quality of the data in \fullname
and indicate its benefits on enhancing LLMs' capability of in-depth code understanding and reasoning
in order to ensure the high accuracy of the generated test cases.
Our improvements over the baseline SantaCoder~\cite{allal2023santacoder} 
on the two languages that it was not trained on (C++ and Go) are on par with the rest.
The advantages are at times even more significant on languages with insufficient training data, e.g., JavaScript. 
This implies that 
it might not be necessary to have our base models 
trained on every language of interest
to generate high-quality tests.
In addition, we observed that our improvements over the top competitors across 
different languages are somewhat dependent on the sufficiency of the tests collected, 
with the least improvement being in JavaScript and the most in Python. 
This observation highlights the contribution of \fullname, 
which is capable of collecting testing data for different languages at scale.

\begin{table*}[t]
\setlength{\tabcolsep}{0.05cm}
\caption{
Evaluations of the completeness of LLM-generated tests.
The ``\#Params'' column indicates the size of models,
and ``\#Pass'' suggests how many tests for $164$ tasks can be executed without any errors.
The ``Line'', ``Stmt'' and ``Branch'' are abbreviations of 
line, statement and branch coverage, respectively.
The models marked with $\dagger$ are intended for test generation.
Limited by the coverage evaluation tools we adopted, branch coverage is not available on C++ and Go.
}
\label{tab:humaneval_cover}
\begin{tabular}{lcccccccccccccccccccc}
\toprule
 &  &  &  & \multicolumn{3}{c}{Python} &  & \multicolumn{2}{c}{C++} &  & \multicolumn{3}{c}{Java} &  & \multicolumn{3}{c}{Javascript} &  & \multicolumn{2}{c}{Go}\\ 
\cmidrule{5-7}\cmidrule{9-10}\cmidrule{12-14}\cmidrule{16-18}\cmidrule{20-21}
\multirow{-2}{*}{Model} & \multicolumn{1}{l}{} & \multirow{-2}{*}{\#Params} &  & \#Pass & Line & Branch &  & \#Pass & Line &  & \#Pass & Line & Branch &  & \#Pass & Line & Branch &  & \#Pass & Stmt\\ 
\midrule
CodeT5p &  & 770M &  & 10.0 & 5.72 & 5.41 &  & 0.7 & 0.43 &  & 40.3 & 4.22 & 2.01 &  & 4.9 & 2.07 & 1.01 &  & 1.7 & 0.73\\ 
CodeGen2 &  & 1B &  & 4.1 & 2.41 & 2.34 &  & 11.6 & 7.07 &  & 52.3 & 5.12 & 3.29 &  & 48.5 & 27.65 & 23.87 &  & 19.2 & 10.99\\ 
WizardCoder &  & 1B &  & 16.1 & 9.39 & 8.95 &  & 3.7 & 2.24 &  & 47.7 & 5.62 & 4.09 &  & 9.2 & 5.50 & 5.32 &  & 0.7 & 0.42\\ 
InCoder &  & 1.3B &  & 3.0 & 1.76 & 1.60 &  & 0.0 & 0.00 &  & 15.0 & 1.54 & 0.91 &  & 0.5 & 0.29 & 0.26 &  & 1.3 & 0.78\\ 
SantaCoder &  & 1.1B &  & 4.5 & 2.62 & 2.59 &  & 4.9 & 2.99 &  & 50.1 & 4.74 & 1.83 &  & 5.9 & 3.53 & 3.23 &  & 0.7 & 0.43\\ 
CAT-LM$^\dagger$ &  & 2.7B &  & 35.9 & 19.51 & 18.03 &  & 0.0 & 0.00 &  & 0.9 & 0.07 & 0.00 &  & 9.2 & 4.53 & 3.49 &  & 0.0 & 0.00\\ 
\textbf{UniTester$^\dagger$ (Ours)} &  & 1.1B &  & 41.2 & 20.71 & 18.27 &  & 28.1 & 13.39 &  & 103.1 & 10.78 & 4.57 &  & 53.3 & 27.59 & 23.34 &  & 36.0 & 12.39\\ 
\bottomrule
\end{tabular}
\end{table*}
\paragraph{RQ.2 How many of the generated tests are complete?}
We presented the average number of passing tests generated by different models and 
the average coverage rates they achieved on the focal functions in Table 3. 
The passing rates of each model across all languages range from $2\%$ to $32\%$. 
Low passing rates indicate the necessity of both accurate and executable test functions generated by LLMs.
Our UniTester demonstrated considerable advantages in this regard, 
with its average passing number being nearly twice that of the top competitor, CodeGen2~\cite{nijkamp2023codegen2}. 
This result shows the inconsistency between the distributions of focal functions and tests, 
emphasizing the necessity of training models with a testing code corpus to synthesize high-quality tests.
We also observed that composing executable tests for C++ was the most challenging among all the languages. 
This is not surprising, as C++ is considered 
relatively weaker in readability and harder to write code in even for human developers.
Moreover, UniTester yielded superior coverage rates, 
with absolute improvements of up to $6.32\%$ and 
an average of $2.8\%$ regarding line coverage. 
Such improvements underscore the effectiveness of UniTester, which was trained on the testing code corpus of \fullname, in covering various conditions in the focal functions.

\paragraph{RQ.3 Is it necessary to train LLMs with pairwise focal and test functions?}
\begin{figure}[t]
    \centering
    \includegraphics[width=\linewidth]{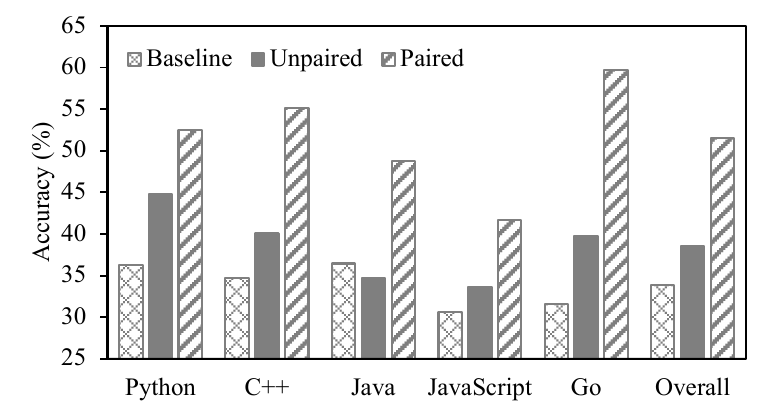}
    \caption{
    Effects of training with pairwise test and focal functions.
    ``Baseline'' is the SantaCoder model, which didn't train on our collected data.
    ``Unpaired'' is the model obtained by training without associating test and focal functions.
    ``Paired'' is UniTester trained with focal-test pairs.
    }
    \label{fig:ablation_pair}
\end{figure}
In \autoref{fig:ablation_pair},
we constructed an ``Unpaired'' variant of UniTester (``Paired'') by 
decoupling test functions from their targets for model training and 
compared it with our models trained with pairwise focal-test data. 
The unpaired variant of UniTester demonstrated its capability to 
generate more accurate tests than the baseline. 
While SantaCoder's paper did not explicitly state the exclusion of testing code 
from their training corpus, 
it is likely that their model was also trained with a certain number of test functions.
However, 
it is shown to be less competent for test generation than the ``Unpaired'' model.
One possible reason for this is the more balanced focal-test functions in \fullname, despite being unpaired.
Furthermore, the remarkable performance advantages of UniTester over 
its unpaired counterpart indicate the importance of associating test functions with their targets. 
Including focal functions in the context when generating tests 
explicitly provides the models with insights into the expected usages and behavior. 
Without these insights, the model can only learn to make reasonable predictions 
for test functions by memorizing all possible focal functions, 
rather than through reasoning. 
This observation highlights the value of \fullname
on not only collecting testing data but also matching it with the focal functions
in a language-agnostic manner.

\paragraph{RQ.4 What are the effects of training with multilingual testing code?}
\begin{figure}[t]
    \centering
    \includegraphics[width=\linewidth]{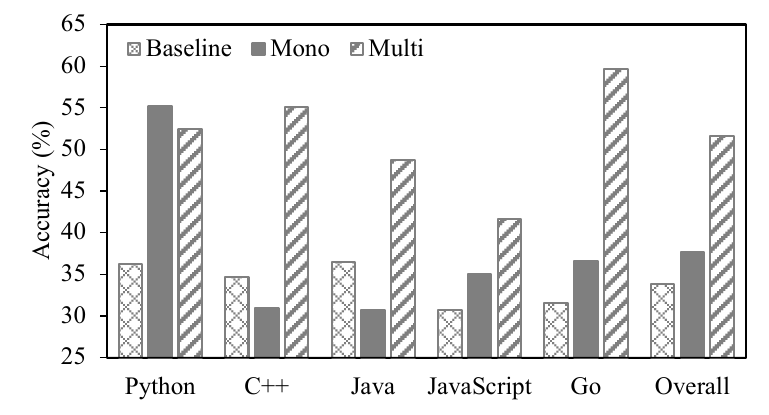}
    \caption{
    Effects of training with multilingual testing code.
    ``Baseline'' is the SantaCoder which didn't train on our collected testing data.
    ``Mono'' and ``Multi`` stand for monolingual and multilingual models trained with either solely Python data or jointly five programming languages.
    }
    \label{fig:ablation_multilingual}
\end{figure}
The construction of all-purpose code generation models
has been garnering increased attention. 
Previous studies have also suggested the potential benefits of 
training on data with shared semantics but different distributions~\cite{feng-etal-2020-codebert}. 
Our proposed \fullname 
is extendable to collect tests for any programming language with an available language server, 
enabling us to build a universal test generation model.
To investigate the best practice of utilizing a multilingual testing code corpus for training LLMs, 
we compared UniTester with its monolingual variant, 
which was trained solely on the Python subset of \fullname. 
We denoted UniTester and its monolingual variant by ``Multi'' and ``Mono''
respectively in \autoref{fig:ablation_multilingual}.
The figure reveals that the monolingual model demonstrated superior capability in 
generating high-quality tests for Python. 
This capability transferred well to other scripting languages like JavaScript and Go, 
but less so to C++ and Java. 
These results suggest modest transferability between syntactically similar languages
but also indicates the potential for negative impacts elsewhere.
Regardless of whether the aim is to build test generation models for 
specific languages or for general purposes, 
a flexible and universal framework for collecting tests from different languages is indispensable. 
This demonstrates the contribution of our proposed \fullname in the field of software testing.

\section{Conclusion}

In this paper, we present \fullname, a novel, diverse, and large-scale dataset 
containing function-level focal-test pairs designed to 
stimulate AI in understanding and writing programs, particularly for test cases. 
This dataset not only excels in size and diversity, but is also effortlessly extendable to other programming languages for specific tasks. 
We further built an autoregressive model based on \fullname to verify the quality of the collected testing code corpus.
This is shown by its superiority in terms of both the accuracy and completeness of the generated tests. 
We hope that \fullname will spur the development of AI for software testing and program understanding in general.

\section*{Limitations}
Despite our effort to design \fullname to involve as little human effort as possible, some manual settings are inevitable.
For example, C++ and JavaScript do not have a commonly used testing framework,
meaning that the developer has to implement the hook to identify test functions for each framework. 
Since we only implemented the test function identification hook for the GoogleTest suite, this accounts for our low repository number in C++.
We believe one can extract more focal-test pairs from the repositories by applying
more precise hooks to our framework.
Our goal is to build an extensible multilingual dataset instead of diving deep into different C++ and JavaScript testing suites.

\section*{Impact Statements}
This paper presents work whose goal is to advance the field of Machine Learning for software testing and program understanding. 
One major impact is that \fullname provides a new dataset to train LLMs.  With paired focal functions and test functions, \fullname is able to advance the quality of the test cases generated by LLMs. In the meantime, we also take the 
generalizability to different programming languages into account and provide an interface where developers can easily extend \fullname to other datasets.
We will release \fullname's code and data upon acceptance.

\bibliography{main}
\bibliographystyle{icml2024}

\newpage
\appendix



\end{document}